\documentclass[%
 reprint,
nofootinbib,
amsfonts,
amsmath,
amssymb,
aps,
prl,
]{revtex4-1}
\usepackage[inner=2cm,outer=2cm, top=2cm, bottom=2cm]{geometry}
\usepackage[english]{babel}
\usepackage[utf8]{inputenc}
\usepackage{amsthm}
\usepackage{mathtools}
\usepackage{physics}
\usepackage{xcolor}
\usepackage{graphicx}
\usepackage{adjustbox}
\usepackage{placeins}
\usepackage[T1]{fontenc}
\usepackage{lipsum}
\usepackage{csquotes}
\usepackage[colorlinks,allcolors=blue!50!black]{hyperref} % For hyperlinks in the PDF
\makeatletter
\renewcommand\frontmatter@abstractwidth{12cm}
\makeatother

\usepackage{dcolumn}
\begin{document}

\title{Materials Graph Transformer predicts the outcomes of inorganic reactions with reliable uncertainties}

\author{Shreshth A. Malik}
    % \email[Correspondence email address: ]{email@institution.com}% Your name
    \affiliation{Cavendish Laboratory, University of Cambridge, Cambridge CB3 0HE, UK.}

\author{Rhys E. A. Goodall}
    % \email[Correspondence email address: ]{email@institution.com}% Your name
    \affiliation{Cavendish Laboratory, University of Cambridge, Cambridge CB3 0HE, UK.}
    
\author{Alpha A. Lee}
    \email[Correspondence email address: ]{aal44@cam.ac.uk}% Your name
    \affiliation{Cavendish Laboratory, University of Cambridge, Cambridge CB3 0HE, UK.}

% \date{\today} % Leave empty to omit a date

\begin{abstract}

A key bottleneck for materials discovery is synthesis. While significant advances have been made in computational materials design, determining synthesis procedures is still often done with trial-and-error. In this work, we develop a method that predicts the major product of solid-state reactions. The key idea is the construction of fixed-length, learned representations of reactions. Precursors are represented as nodes on a `reaction graph', and message-passing operations between nodes are used to embody the interactions between precursors in the reaction mixture. We show that this framework not only outperforms less physically-motivated baseline methods but also more reliably assesses the uncertainty in its predictions. Moreover, our approach establishes a quantitative metric for inorganic reaction similarity, allowing the user to explain model predictions and retrieve relevant literature sources.

\end{abstract}

\maketitle

%BG/wider field - Materials Discovery
The discovery of new materials and control of their properties is fundamental to the advancement of technology. Adjusting the structure and composition of crystals allows fine-tuning of properties such as conductivity and band gap \cite{DInnocenzo:2014aa, Kalantari2011, Gong:2014aa}. Advances in first-principles calculations and increased computational power have revolutionised the availability of materials information \cite{MatProj2013, OQMD}. Simultaneously, advances in computational materials design \cite{Jansen:2015aa} have enabled high-throughput prediction and screening in silico \cite{Ludwig:2019aa}.

%Subfield and Gap - Synthesis
However, due to the challenges associated with material synthesis, only a small proportion of candidates are produced in the laboratory and experimentally validated. The most common route to the formation of polycrystalline materials is solid-state synthesis \cite{Kanatzidis:2017aa}, which typically involves calcining a mixture of solid reactants. Reactions occur through solid-state diffusion of ions, thus reactants are often milled and mixed to improve the reaction kinetics. However, the interplay between thermodynamically driven energy minimisation and kinetic factors results in products which are often difficult to predict; the field lacks the well-understood reaction mechanisms in organic chemistry. Particular procedures or reactants result in the formation of crystals with unique morphologies and compositions as they form through metastable states. Thus researchers frequently rely on a trial-and-error process of synthesis and evaluation. The development of tools to propose and evaluate the most promising synthetic pathways is, therefore, one of the biggest challenges facing materials discovery \cite{Aykol:2019aa, kim2019machinelearned}.

%Related Work and Limitations 
A recently released inorganic synthesis dataset has now enabled the possibility of using a data-driven approach to predict the outcomes of inorganic synthesis \cite{Kononova:2019aa}. To our knowledge, there exists only one study which explores this opportunity \cite{kim2018inorganic}, which addresses the inverse problem of retrosynthesis -- predicting precursors that can react to yield a target product -- with an architecture that cannot be readily extended to solve the forward problem. However, save for large scale experimental validation, the accuracy of retrosynthesis models cannot be quantitatively benchmarked because there are almost infinitely many ways to synthesize a material, and reactions reported in the literature are not necessarily even the best synthetic route.  

Here, we report an accurate forward reaction prediction model with reliable uncertainties. Methodologically, we leverage ideas from graph representation learning to learn the optimal representation of sets of inorganic reactants directly from data.  Our work significantly outperforms baselines, and we show that model uncertainty can be robustly estimated using an ensemble.
% Moreover, our method can provide explainability to predictions of novel reactions by assessing the similarity of the reaction representation to known reactions.

Our work is a building block towards an inorganic retrosynthesis planner that can interface directly with computational tools to close the materials design-make-test cycle.

\section{Data and Model}
\subsection{Solid-State Syntheses Dataset}
\label{sec:data}

We can define a generalised solid-state synthesis procedure for a target material as a sequence of processes (\textit{actions}) performed on a set of starting materials (\textit{precursors}). A recent study has extracted detailed information for over 19,000 such synthesis procedures from academic literature \cite{Kononova:2019aa}. Each synthesis procedure has four relevant fields for this problem,

\begin{enumerate}
    \item \textbf{Target:} The stoichiometric formula of the target material for the synthesis.
    \item \textbf{Precursors:} Defined as starting materials which explicitly share at least one element with the target material, excluding `abundant' materials, i.e. those found in the air.
    \item \textbf{Processing Actions:}  The sequence of synthesis actions performed on the precursors, including the relevant conditions for each action where available.
    \item \textbf{Balanced Chemical Equation:} Balanced chemical equations for the formation of the target compound provide the relevant molar ratios of precursors in the reaction mixture.
\end{enumerate}

\noindent
This dataset is to our knowledge the most extensive and highest-fidelity available for inorganic syntheses. Chemical information (precursors, targets and balanced equations) has been extracted with high accuracy (93\%). 

However, there are some notable limitations that must be considered when utilising it for modelling. Firstly, some procedures have actions and conditions with missing or incorrect data. Over 10\% of reactions have no, or only one recorded action. This is primarily because standard procedures are often referenced rather than explicitly stated. The accuracy of the complete procedures (where all actions, conditions, and chemical information are fully correct) is 51\%. Further, there is no structural information for the products. This limits physical analysis of the materials and hence a purely stoichiometric approach must be used in this work. 

Nevertheless, this dataset provides a wealth of structured information on successful solid-state syntheses. Machine learning (ML) frameworks provide a unique way to assimilate empirical patterns from data. One can thus envisage the target material as being an abstract but learnable function of the precursors and processing procedure. 

\begin{figure*}
    \centering
    \includegraphics[width=1\textwidth]{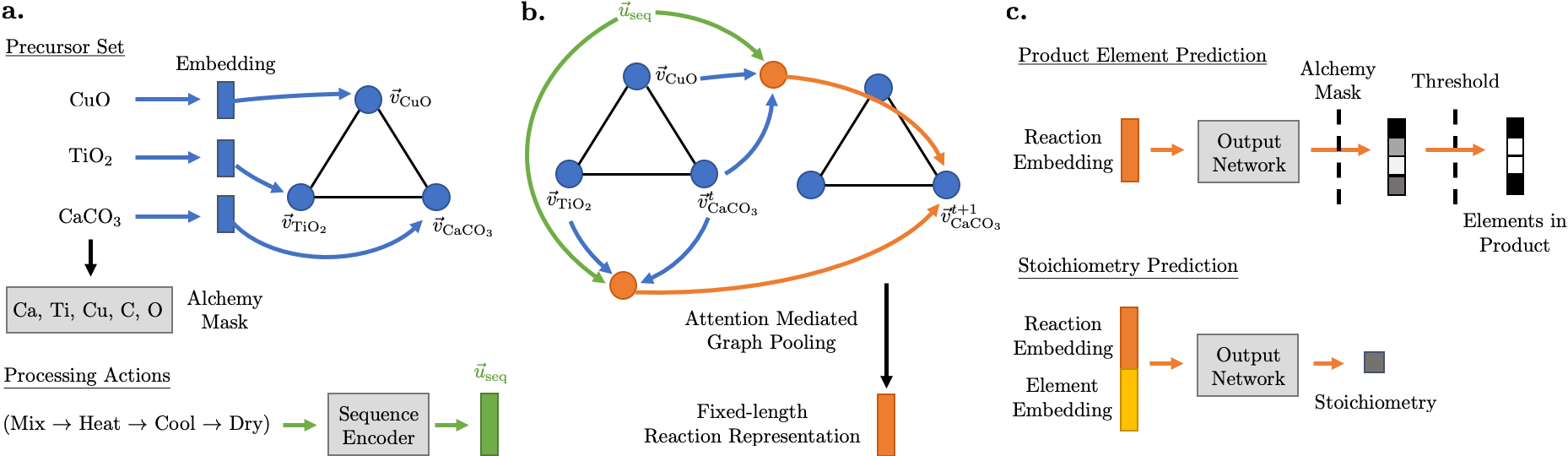}
    \caption{
    \textbf{a.} Representation of precursors and processing actions. \textit{Magpie} embeddings \cite{Ward:2016aa} are used to represent the precursors as fixed-length vectors. These are used as features for nodes on a dense graph. An \textit{alchemy} mask, used to constrain the predictions of the model, is constructed from the set of elements in the precursors. Processing action sequences are encoded using an LSTM encoder. \textbf{b.} Reaction representation learning. The reaction graph goes through a series of message-passing operations between nodes. The action embedding is used as a global state in all the message-passing steps. The final graph representation is then pooled, weighted by attention coefficients, to obtain a fixed-length representation of the reaction. \textbf{c.} Product prediction from reaction representations. The learned reaction embedding is used to predict the elements present in the major product. The \textit{alchemy mask} ensures that only elements present in the precursors can be predicted as present in the product. Separately, the amount of each predicted element is regressed with its fractional stoichiometry. \textit{Matscholar} element embeddings \cite{Tshitoyan:2019aa} are concatenated with the reaction embedding to query the element in question.} \label{fig:model}
\end{figure*}

\subsection{Representation of Inorganic Reactions}
\label{sec:rep-mat}

ML models generally operate on fixed-length inputs whilst typically our understanding of inorganic reactions is expressed in terms of variable-sized sets of precursors and sequences of processing actions. Bridging this discrepancy is a key challenge that needs to be addressed in the construction of ML algorithms for inorganic reaction prediction. 

Here this process is broken down into 2 stages: 1) obtaining fixed-length representations for the precursors and the processing sequence, and 2) performing set regression on this set of precursors to predict the product given the processing sequence.

\noindent
\subsubsection{Representation of Precursors} \vspace{0.2cm}

\noindent    
A na\"{i}ve approach would consider representing materials as sparse vectors with components proportional to the relative amounts of its constituent elements (its stoichiometry). However, such a representation fails to capture critical correlations between different elements. In contrast, \textit{Magpie} embeddings offer powerful general-purpose descriptors for bulk inorganic materials in the absence of structural information \cite{Ward:2016aa}. These are highly engineered, 145-dimensional vector representations which include as many scientific priors derived from its chemical formula as possible. These features fall into 4 groups: stoichiometric properties, elemental properties, electronic structure, and ionic compound features. We utilise these embeddings as initial precursor feature vectors.

\subsubsection{Representation of Processing Actions}
\label{sec:rep-seq}

Sequence representation is a well studied problem in ML research \cite{seq2seq,vaswani2017attention, devlin2018bert}. Here we utilise single-layer Long Short-Term Memory Units (LSTMs) \cite{lstm} in an autoencoder architecture to obtain fixed-length representations of the processing sequences.

\subsection{Reaction Graph Model}
\label{sec:model}

In solid-state syntheses, precursors are inherently `mixed' together to form a product. Here we make use of an architecture for set regression based on the \textit{Roost} model introduced in \cite{goodall2019predicting}. The model treats this process through a series of message-passing stages in which precursors appear as nodes on a dense, weighted graph. A fixed-length representation for the reaction is derived from this graph which allows for generalisation to arbitrary numbers of precursors. 

This learned reaction embedding is then used to predict the elements present in the target material. The same embedding is then used to separately predict the product stoichiometry given the elements in the product. If the stoichiometry was directly predicted, an arbitrary threshold stoichiometry would have to be used to select which precursor elements are actually present in the product. The two stage approach used here is therefore necessary to differentiate between products with and without trace elements. 

\subsubsection{Reaction Representation Learning}        
\label{sec:model-comp}

The crucial first step is to produce a fixed-length representation of the reaction mixture. Figure \ref{fig:model}\textbf{a} shows how precursors are represented as nodes on a dense, weighted graph and action sequences are transformed into a fixed-length representation using an LSTM encoder. The initial precursor feature vectors are transformed into a trainable embedding through a learnable affine transformation,
\begin{equation}
\label{eq:embedding}
\vec{v}_i = \textbf{W}\vec{v}_{0,i} ,
\end{equation}
where $\textbf{W}$ is a learnable weight matrix, and $\vec{v}_{0,i}$ is the initial precursor representation.

A series of message-passing operations (Figure \ref{fig:model}\textbf{b}) then update the precursor representations
\begin{equation}
    \vec{v}_{i}^{t+1} = U^{t}(\vec{v}_{i}^{t}, \sum\nolimits_{j\neq i}\vec{v}_{j}^{t}, \vec{u}_{seq}),
\end{equation}
where $\vec{v}_i^{t}$ is the $i$th precursor feature vector at message-passing layer $t$, $\vec{v}_j^{t}$ are the other precursor features, $\vec{u}_{seq}$ is the fixed-length representations of the action sequences and $U^{t}$ is the update function.

At the heart of the update function is the soft-attention mechanism, which allows the model to capture the importance of the other precursors in the reaction to the precursor in question. \textit{Attention} is in general a method for assigning the importance of certain features in a machine learning task. A \textit{soft-attention} mechanism extends this idea by allowing the model to learn the attention coefficients itself \cite{vaswani2017attention}. This has been shown to improve graph representation learning \cite{velikovi2017graph}. Here, unnormalised attention coefficients are first calculated across pairs of precursors
\begin{equation}
\epsilon_{ij} = f\left( \vec{v}_{i}^{t} \oplus \vec{v}_{j}^{t} \oplus \vec{u}_{seq}\right),
\end{equation}
where $f$ is a single hidden-layer neural network, $\oplus$ is the concatenation operation, and $\vec{u}_{seq}$ is the action sequence embedding which is used as a `global' state in all update steps. This gives procedural context to enable differentiation between products formed through different synthesis pathways. $\epsilon_{ij}$ are then normalised by a weighted softmax function,

\begin{equation}
\label{eq:weighted-softmax}
\alpha_{ij} = \frac{\displaystyle w_{ij} \exp(\epsilon_{ij})}{ \displaystyle \sum\nolimits_{k} w_{ik} \exp(\epsilon_{ik})},
\end{equation}
where the weight $w_{ij}$ is the molar amount of precursor $j$ in the balanced chemical equation. This gives the model context of the quantities of each precursor in the reaction mixture. The node feature is then updated with the pairwise interactions between nodes, weighted by their attention coefficients
\begin{equation}
\vec{v}_{i}^{t+1}= \vec{v}_{i}^{t} + \frac{1}{K} \sum_{k=1}^{K} \sum_{j\neq i}\alpha_{ij}^{k}g^{k}\left( \vec{v}_{i}^{t} \oplus \vec{v}_{j}^{t} \oplus \vec{u}_{seq}\right).
\end{equation}     
Here $g$ is again a single hidden-layer neural network, and we average over $K$ attention heads, which has been shown to stabilise performance \cite{velikovi2017graph}.

A similar attention-mediated pooling operation is then used to create a fixed-length, learned representation of the reaction $\vec{r}$ from the final graph. The model learns how much attention to pay to each precursor given its final representation and relative quantity.

\subsubsection{Product Prediction}

The learned reaction embedding is then passed through a feed-forward neural network trained to perform multi-label element classification for elements in the target product of the reaction (Figure \ref{fig:model}\textbf{c}). In materials synthesis, the products cannot contain elements that are not present in the precursors. We add this inductive bias in the form of a binary \textit{`alchemy mask'} on the output of the network. Oxygen is assumed to be abundant and is additionally included in all masks irrespective of its presence in the precursors. 
%Other gases in the air are in general too inert for involvement in solid-state reactions.

Once the elements in the product have been predicted, the final task is to predict the relative amounts of each element in the product. Here, we use \textit{Matscholar} element embeddings, $\vec{e}_i$, to represent and differentiate between constituent elements \cite{Tshitoyan:2019aa}. These are concatenated with the reaction representation, $\vec{r}$, to provide a contextualised query and fed through another feed-forward neural network, $h$, to predict the relative amount of element $i$,
\begin{equation}
a_i=  h\left( \vec{e}_i \oplus \vec{r} \right).
\end{equation}
These are then normalised using a softmax function to arrive at a fractional stoichiometry
\begin{equation}
s_i= \frac{1}{N} \sum_{k=1}^{N} \text{softmax}_{i}(a^k_i),
\end{equation}     
where $s_i$ is the normalised stoichiometry of element $i$ in the product, and we average over $N$ predictions to increase stability and accuracy. In addition, using an ensemble of $N$ models also allow us to estimate the epistemic uncertainty by taking the variance in predictions across the models in the ensemble. We refer to this ensembled model as the reaction graph model.

\section{Results and Discussion}
\label{sec:results}

\subsection{Product Prediction}
\label{sec:results-main}

\begin{figure}[t]
    \centering
    \includegraphics[width=0.98\linewidth]{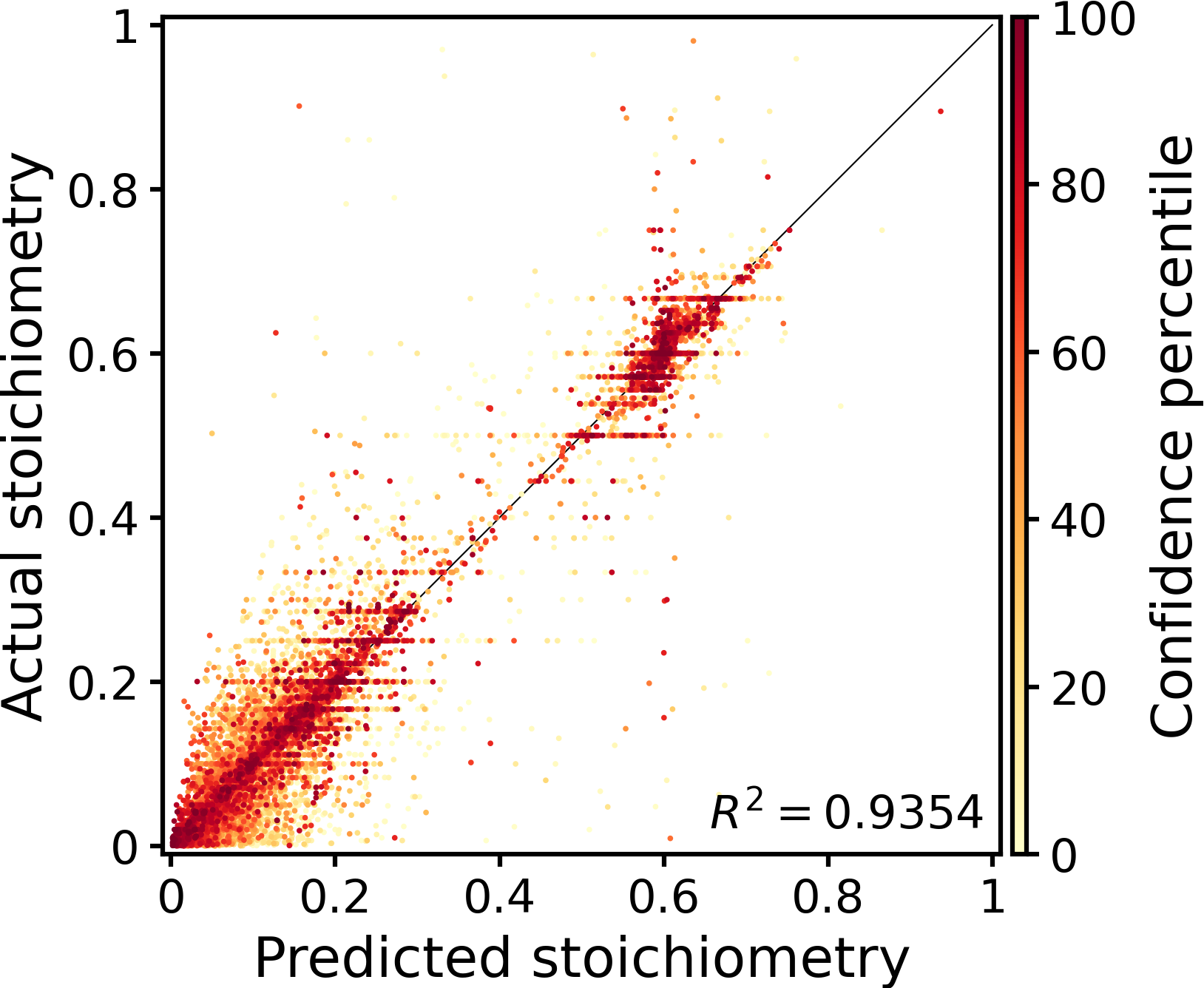}
    \caption{Parity plot for product stoichiometry prediction. The fractional amount of each element in the target product is plotted against its predicted value. The data points are coloured according to their relative confidence, estimated from the variation in predictions from the ensemble. In general, more confident predictions appear closer to the equivalence line.}
    \label{fig:stoich}
\end{figure}

The reaction graph model was tested on solid-state reactions with up to ten precursors.  Product elemental composition was predicted with a subset accuracy of 0.940. This can be compared to a null baseline accuracy of 0.338 which would be achieved if all elements in the precursors were assumed to be present in the product. There is evidently a bias towards positive element labelling; 83.9\% of elements in the precursors are also present in the product. Thus another metric of interest is the F1 score. The average F1 score achieved was 0.9910, where the score for each element is weighted by the number of true occurrences to account for the elemental imbalance.

Results for stoichiometry prediction can be seen in Figure~\ref{fig:stoich}. The accuracy of model predictions is well aligned with the epistemic uncertainty -- the most certain predictions lie closest to the equivalence line. The L1 and L2 distances between composition vectors can provide metrics for measuring material similarity. Here, the fractional stoichiometry vectors of predicted products were accurate to a mean L1 distance of 0.1259, and L2 distance of 0.0729 to their true stoichiometries.

\subsection{Explaining Predictions via Reaction Similarity}
\label{sec:similarities}

The predictions of the model can be explained by considering the reaction representation -- similar feature vectors will result in similar predictions. We can therefore understand  why certain predictions are made through assessing the cosine similarity between the reaction embedding vector and those in the training data. We illustrate this via examples of such reactions that are not in the training set, and show that similar reactions in the training set provide explanations for these predictions. 

Firstly we investigate the synthesis of a sodium-ion battery cathode material NVP \cite{GOVER20061495},
\begin{equation}
\label{eq:vpo4}
    \text{3 NaF + 2 VPO\textsubscript{4}} \rightarrow \text{Na\textsubscript{3}V\textsubscript{2}(PO\textsubscript{4})\textsubscript{2}F\textsubscript{3}.}
\end{equation}
The product composition was predicted accurately within an L2 error of 0.0067. To analyse why the prediction was accurate, we assess the similarity of the learned reaction embedding derived from the precursors and processing action sequence to other reactions in the training set:
\begin{table}[!h]
\centering
\label{tab:vpo4}
\begin{tabular}{p{6.5cm} p{1.5cm}}
\textbf{Nearest Neighbors to (\ref{eq:vpo4})} & \textbf{Similarity} \\
\hline
LiF + VPO\textsubscript{4} $\rightarrow$ LiVPO\textsubscript{4}F  & 98.5\% \\ \hline
LiF + FePO\textsubscript{4} $\rightarrow$ LiFePO\textsubscript{4}F & 97.3\% \\ \hline
\end{tabular}
\end{table}

Here we see that the embedding is very similar to other vanadium fluorophosphate syntheses. Despite differing cations appearing in the precursors, the model has learnt the general form of such reactions and the expected ratios of elements in the product. The model triangulates a prediction from similar observed reactions and the elemental context.  

As a second example, we explore a different subspace of materials. Perovskite manganites with the general formula Ln\textsubscript{1-x}A\textsubscript{x}MnO\textsubscript{3} (where Ln is a trivalent rare earth and A is a divalent alkaline earth) have attracted interest since the discovery of large negative magneto-resistance in such structures \cite{magnetores}. When faced with the novel reaction \cite{DEBBEBI201867},
\begin{multline}
\label{eq:managanite}
       \text{0.05 MnO\textsubscript{2} + 0.15 CaCO\textsubscript{3} + 0.3 La\textsubscript{2}O\textsubscript{3} +} \\ \text{0.01 BaCO\textsubscript{3}} \rightarrow \text{La\textsubscript{0.6}Ca\textsubscript{0.15$\cdot$ 0.05}Ba\textsubscript{0.2}MnO\textsubscript{3} +} \\
       \text{0.16 CO\textsubscript{2} + 0.505 O\textsubscript{2},}
\end{multline}
the product was predicted within an L2 error of 0.0019. The reaction embedding was observed to be similar to other manganite syntheses in the training set:

\begin{table}[!h]
\centering
\label{tab:managanite}
\begin{tabular}{p{6.3cm} p{1.7cm}}
\textbf{Nearest Neighbors to (\ref{eq:managanite})} & \textbf{Similarity} \\
\hline
0.083 Pr\textsubscript{6}O\textsubscript{11} + 0.47 CaCO\textsubscript{3} + 0.03 BaCO\textsubscript{3} \hspace{0.2cm} +  MnO\textsubscript{2} $\rightarrow$ Pr\textsubscript{0.5}Ca\textsubscript{0.47}Ba\textsubscript{0.03}MnO\textsubscript{3} \hspace{1cm} + 0.5 CO\textsubscript{2} + 0.208 O\textsubscript{2} & 97.6\% \\ \hline
0.05 SrCO\textsubscript{3} + 0.15 CaCO\textsubscript{3} + 0.5 Mn\textsubscript{2}O\textsubscript{3} \hspace{0.5cm} + 0.05 BaCO\textsubscript{3} + 0.375 La\textsubscript{2}O\textsubscript{3} + 0.063 O\textsubscript{2} $\rightarrow$ La\textsubscript{0.75}Ca\textsubscript{0.15}Sr\textsubscript{0.05}Ba\textsubscript{0.05}MnO\textsubscript{3} + 0.25 CO\textsubscript{2} & 97.5\%
\\ \hline
\end{tabular}
\end{table}
The model predicted the correct ratios of oxygen and manganese to the other elements in the product, learning from ratios in similar reactions. In the dataset however, the product composition was incorrectly extracted as a mixture of La\textsubscript{0.6}Ca\textsubscript{0.15} and Ba\textsubscript{0.2}MnO\textsubscript{3} in the ratio 1:0.05 due to the abnormal dot notation in the calcium stoichiometry. By comparison to similar reactions, one can thus also validate reactions extracted from literature.

The reaction embeddings also serve as a means to analyse why certain predictions are incorrect. As many of these failure modes can be identified via estimation of the epistemic uncertainty, we are primarily interested in explaining the worst predictions -- those which have large error but low uncertainty. Reaction (\ref{eq:NaCoO2}) is an example of one such case.
\begin{multline}
\label{eq:NaCoO2}
       \text{1.5 Na\textsubscript{2}CO\textsubscript{3} + 0.333 Co\textsubscript{3}O\textsubscript{4}} \rightarrow \\ \text{Na\textsubscript{3}CoO\textsubscript{2} + 1.5 CO\textsubscript{2} + 0.417 O\textsubscript{2}}
\end{multline}
\begin{table}[!h]
\centering
\label{tab:over-ox}
\begin{tabular}{p{2cm}p{6cm}}
\hline
\textbf{Target}  & Na: 0.5000 \hspace{0.5cm}   Co: 0.1667  \hspace{0.5cm}    O: 0.3333    \\ \hline
\textbf{Prediction} &  Na: 0.1985 \hspace{0.5cm}   Co: 0.2692 \hspace{0.5cm}    O: 0.5323 \\ \hline
\end{tabular}
\end{table}

In the original paper \cite{ZHOU2006338}, Na\textsubscript{3}CoO\textsubscript{2} is the nominal product but the paper actually reports the formation of different phases of Na\textsubscript{x}CoO\textsubscript{2} in their experiment. As such, the model prediction is not unreasonable. The model reached that prediction by inferring from similar reactions observed in the training data (but not in ref \cite{ZHOU2006338}):
\begin{table}[!h]
\centering
\label{tab:NaCoO2}
\begin{tabular}{p{6.3cm} p{1.7cm}}
\textbf{Nearest Neighbors to (\ref{eq:NaCoO2})} & \textbf{Similarity} \\
\hline
0.375 Na\textsubscript{2}CO\textsubscript{3} + 0.333 Co\textsubscript{3}O\textsubscript{4} + 0.146 O\textsubscript{2} $\rightarrow$ Na\textsubscript{0.75}CoO\textsubscript{2} + 0.375 CO\textsubscript{2} & 99.9\% \\ \hline
0.5 Na\textsubscript{2}CO\textsubscript{3} + 0.333 Co\textsubscript{3}O\textsubscript{4} + 0.083 O\textsubscript{2} $\rightarrow$ NaCoO\textsubscript{2} + 0.5 CO\textsubscript{2} & 99.9\%
\\ \hline
\end{tabular}
\end{table}
\subsection{Benchmarking and Ablation Study}
\label{sec:ablation}

\begin{figure*}
    \centering
    \includegraphics[width=1\linewidth]{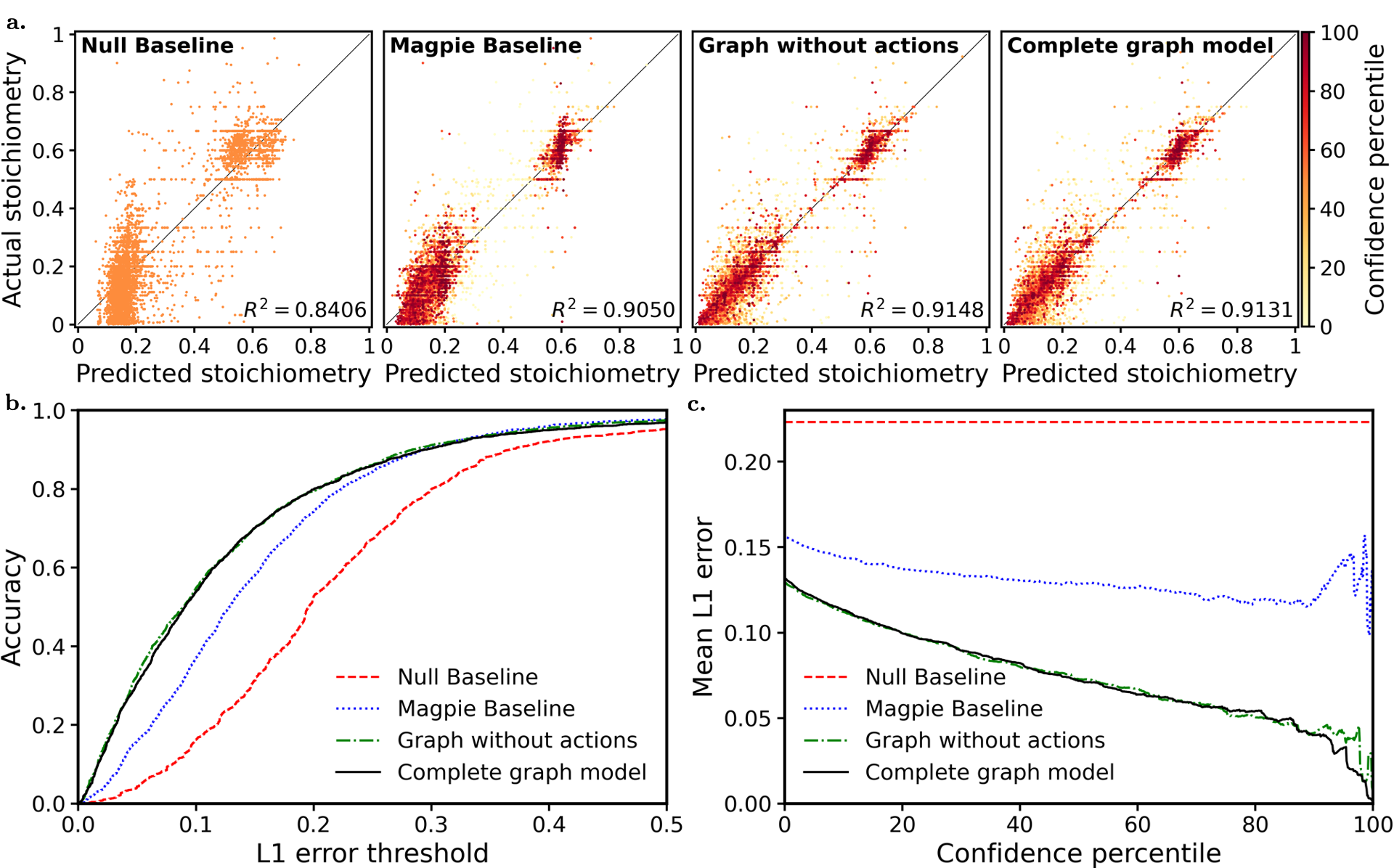}
    \caption{\textbf{a.} Ablation study parity plots for product stoichiometry prediction. An improvement in both correlation and uncertainty estimation is seen in incorporating the reaction graph framework. \textbf{b.} Threshold plot for product stoichiometry prediction. The cumulative proportion of correctly predicted stoichiometries is plotted against an L1 error threshold. A clear improvement is seen through using the learned reaction graph embedding compared to fixed \textit{Magpie} features. No apparent improvement is observed through the inclusion of the action sequences. \textbf{c.} Confidence-error plots for stoichiometry prediction showing the mean L1 error in product stoichiometry prediction as the most uncertain predictions are removed. Uncertainty estimation is greatly improved through using the reaction graph embedding -- more erroneous predictions are in general more uncertain.} \label{fig:ablation}
\end{figure*}

To place the accuracy of our model into context, we benchmark our reaction graph model against a baseline approach and conduct an ablation study on its key features. 

The baseline model we use takes the \textit{Magpie} precursor embeddings and concatenates them to be used as the input to a neural network with the same architecture as the output network of the reaction graph model. As with the reaction graph model, we make use of an ensemble of $N$ individual models for this baseline.

While the reaction graph model can handle variable numbers of precursors, this \textit{Magpie} baseline approach is inherently restricted to a certain input size. Thus we limit our model to reactions with up to three precursors only, and zero-pad the input in the \textit{Magpie} baseline model where there are less than three precursors to enable comparison.

The ablation study compares results across reaction graph models with/without the processing actions and the \textit{alchemy mask} to discern which parts of the model design are significant. 
The results from the study are shown in Figure \ref{fig:ablation} and a summary of the key accuracy metrics is shown in Table~\ref{tab:elempred}. A null baseline result is also given for comparison where we assume all elements in the precursors are present in the product, and are present in their average relative amounts derived from the training data.

\subsubsection{Reaction Representation Learning}

For product element prediction, the reaction graph models are seen to perform significantly better than using \textit{Magpie} features alone. For stoichiometry prediction, Figure~\ref{fig:ablation}\textbf{b} shows the proportion of predictions that are within given L1 error tolerances. A clear improvement is seen through incorporating the graph representation learning framework. 

Figure \ref{fig:ablation}\textbf{a} shows plots of actual against predicted fractional stoichiometry. This figure shows how predictions are distributed about the equivalence line, and their relative confidence as measured through ensemble variance. Similar to the null baseline, the \textit{Magpie} baseline model predictions show a lack of parity about the equivalence line. 
%implying that the model has not effectively learnt how different reaction mixtures affect product stoichiometry.

In contrast, the learned reaction graph embeddings show a significant improvement, both in absolute terms and in better prediction of the model uncertainty. In general, the most certain predictions are closest to the equivalence line and there is more parity. This indicates the model has assimilated some chemical knowledge. The improvement in uncertainty prediction can be seen more clearly in the confidence-error plot in Figure \ref{fig:ablation}\textbf{c}. Including the learned reaction graph embedding shows a clear decrease in the mean error of predicted products as the most uncertain predictions are removed. The almost flat line observed when using fixed \textit{Magpie} embeddings indicates that there is little calibration between error and uncertainty for the baseline model.

\subsubsection{Alchemy Mask}
The results in the lower half of Table \ref{tab:elempred} are achieved through the inclusion of the \textit{alchemy mask}, which effectively prevents the model from making predictions which are chemically unfeasible. This inductive bias provides the most notable increase in accuracy across the board %(+34\% on average) 
for element prediction. This highlights the utility of incorporating physical principles into ML models.

\begin{table*}
\centering
\caption{Ablation performance table for reactions with up to three precursors. The subset accuracy and weighted F1 score is shown for the prediction of product elements. The null baseline result assumes all elements in the precursors are present in the product, and assumes that elements are present in their average amounts from the training data. Providing inductive bias to the model in the form of the alchemy mask improves accuracies across the board. The reaction graph framework shows a clear improvement over the \textit{Magpie} baseline. Removing the action context does not have a notable effect on model accuracy. \\}
\label{tab:elempred}
\begin{tabular}{lcccccc}
% \hline
\toprule
Reaction Representation & Alchemy & Subset & F1 Score & Mean & Mean & $R^2$ \\ 
% Stoichiometry &  &  0.815(5) & 0.967(3) &  &  & \\
 & Mask & Accuracy &  & L1  & L2 & \\ \hline
\textit{Magpie} Baseline &  &  0.625 & 0.902 &  &  & \\
Reaction Graph without Actions &  &  0.773 & 0.950 &  &  & \\
Complete Reaction Graph & & 0.770 & 0.948 &  &  & \\ 
\hline
Null Baseline  & $\checkmark$ & 0.416   & 0.988 & 0.2233 & 0.1333 & 0.8406 \\
\textit{Magpie} Baseline & $\checkmark$ & 0.933 & 0.990 & 0.1568 & 0.0948 & 0.9050 \\
Reaction Graph without Actions & $\checkmark$ & 0.960 & 0.994 & 0.1298 & 0.0790 & 0.9148 
 \\
Complete Reaction Graph & $\checkmark$ & 0.960 & 0.994 & 0.1326 & 0.0805 & 0.9131 \\ \hline
\end{tabular}
\end{table*}

\subsubsection{Processing Action Sequences}
\label{sec:actions}

The synthesis actions are crucial in determining the major product of an inorganic reaction; different procedures on the same precursors can lead to differing products due to the interplay between thermodynamics and kinetics. However, Table~\ref{tab:elempred} shows the addition of the action sequences in the reaction embedding has a detrimental effect -- if any -- on model performance. 

The level of detail of the action data used in the model can provide an explanation. Different product compositions and morphologies can be formed from even slight changes in temperature or pressure. Thus one cannot expect that using the generic, coarse terms to represent action sequences can lead to significant improvements in performance. The inaccuracies of the processing sequences in the dataset only serve to compound this effect. Thus a model that includes more granular processing procedures is required for practical applications. 

Despite this, when the models were extended to the full dataset by including reactions with up to ten precursors, there was a significant improvement in accuracy through incorporating the processing actions sequences. An L1 error of 0.1357 was achieved without the actions, and an L1 error of 0.1259 with the actions. This is likely due to the greater diversity of reactions in the extended dataset, meaning the actions are more important in providing contextual information.

\section{Conclusion}

In this work, a data-driven approach to inorganic reaction prediction has been investigated for the first time. A physically-motivated model has been developed which can predict major products of solid-state reactions from precursors and synthesis procedures. The key feature of this method is the generation of a learned reaction embedding which effectively encodes the reaction context for product prediction. Through an ablation study, this framework is shown to predict product stoichiometries more accurately than less physically-motivated models. Importantly, it also more reliably assesses the uncertainty in its predictions.
    
The learned reaction representations have been shown to provide explainability to novel predictions through analysing similarities with known reactions in the literature. These embeddings can also be transferred to further problems. One can envisage using it to predict the probability of success of a given synthesis procedure. This would further aid retrosynthesis; precursors and synthesis pathways which have the highest probability of success can be preferentially chosen. This would require supervision in the form of both successful and unsuccessful reactions \cite{Raccuglia:2016aa}. 

This work provides a basis for evaluating a future retrosynthesis planner for inorganic materials. For practical use, however, further work is required to differentiate between morphologies of products and more granular synthesis action steps with relevant conditions are needed. Access to greater quality and quantity of synthesis data and a concerted effort between ML practitioners and experimentalists will allow these advances.

\section*{Methods}

\subsection{Data Processing}
\label{sec:comp-data}

Reactions with non-stoichiometric or organic precursors and targets, and reactions with pure products or only one precursor were removed from the dataset. This reduced the dataset size to 16,231 reactions with up to ten precursors, and 11,083 reactions for the ablation study with up to three precursors. This was randomly split into training and test sets with an 80:20 ratio.

Processing action sequences ranged from 0 to 16 steps in length, and were represented by 7 unique action types: \{Dry Mixing, Mixing in solution, Quenching, Shaping, Heating, Drying, Liquid Grinding\}.

Precursor stoichiometries were extracted along with their molar ratios from the balanced chemical equation for each reaction. Where precursors or targets were themselves a mixture of different materials/phases, their stoichiometries were added, weighted by their amounts. 

Targets were extracted as 81-dimensional stoichiometric vectors. The set of elements present in the precursors was used to construct the \textit{alchemy mask}.

\subsection{Implementation Details}
\label{sec:comp-imp}

The model was implemented in PyTorch~\cite{NEURIPS2019_9015} and \textit{Matminer} was used for generating \textit{Magpie} features from precursor stoichiometries \cite{matminer}. 

% Long Short-Term Memory Units (LSTMs) were used as recurrent units for the action auto-encoder, which handle the vanishing gradient problems encountered in traditional RNNs \cite{lstm}. 
An embedding dimension of 8 and a hidden state dimension of 32 were used for the processing actions autoencoder. This lead to a sequence reconstruction accuracy of $\sim87\%$. Reconstruction accuracies of over 95\% were achieved using a hidden dimension of 64 but did not significantly improve product prediction accuracy.

The learnable precursor embedding size was set to 128, and 5 message-passing layers were used, each with 3 attention heads. The hidden layer dimension was chosen to be 256 for both $f$ and $g$. The output network of the product element prediction model had hidden layer dimensions of 256, 512, 512 and 256 respectively with ReLU activation. The threshold output probability for elemental presence was chosen through maximising accuracy on a validation set. The stoichiometry prediction model had hidden layer dimensions of 256, 256, 256, 256, 128, 128 and 64 respectively with ReLU activation. These were averaged over 5 iterations. Skip connections were used to reduce the vanishing gradient effect \cite{resnet}. 

Learning rates were chosen using a heuristic search where we choose a rate one magnitude lower than that which causes the training loss to diverge. A mini-batch size of 256 and the Adam optimiser was used to train the model \cite{kingma2014adam}. Early stopping using an 80:20 train:validation split was used in all models to avoid over-fitting. 
 
The ensembles used to estimate the epistemic uncertainty in both the \textit{Magpie} baseline and reaction graph models were constructed using 5 individual models trained on the same data with different initialisation. 

%This lack of structural information is a major limitation of the dataset as it limits material understanding. A material is defined not only by its stoichiometry but crucially by its structure. The dataset and therefore this framework is completely structure-agnostic and is thus limited in its predictive abilities. The use of predicted crystal structures from stoichiometry could be an avenue of further investigation \cite{Zhao:2020aa}.

\subsection*{Data availability}
The exact data used in this work can be found at \url{https://doi.org/10.6084/m9.figshare.9722159.v3}. More recent versions of the synthesis dataset are released by the original authors at \url{https://github.com/CederGroupHub/text-mined-synthesis_public}. 
\subsection*{Code availability}
An open-source repository of the code used in this work is available at \url{https://github.com/s-a-malik/inorg-synth-graph}.

\section*{Acknowledgements}
R.E.A.G. and A.A.L. acknowledge the support of the Winton Programme for the Physics of Sustainability.

\bibliography{references.bib}

\begin{thebibliography}{29}%
\makeatletter
\providecommand \@ifxundefined [1]{%
 \@ifx{#1\undefined}
}%
\providecommand \@ifnum [1]{%
 \ifnum #1\expandafter \@firstoftwo
 \else \expandafter \@secondoftwo
 \fi
}%
\providecommand \@ifx [1]{%
 \ifx #1\expandafter \@firstoftwo
 \else \expandafter \@secondoftwo
 \fi
}%
\providecommand \natexlab [1]{#1}%
\providecommand \enquote  [1]{``#1''}%
\providecommand \bibnamefont  [1]{#1}%
\providecommand \bibfnamefont [1]{#1}%
\providecommand \citenamefont [1]{#1}%
\providecommand \href@noop [0]{\@secondoftwo}%
\providecommand \href [0]{\begingroup \@sanitize@url \@href}%
\providecommand \@href[1]{\@@startlink{#1}\@@href}%
\providecommand \@@href[1]{\endgroup#1\@@endlink}%
\providecommand \@sanitize@url [0]{\catcode `\\12\catcode `\$12\catcode
  `\&12\catcode `\#12\catcode `\^12\catcode `\_12\catcode `\%12\relax}%
\providecommand \@@startlink[1]{}%
\providecommand \@@endlink[0]{}%
\providecommand \url  [0]{\begingroup\@sanitize@url \@url }%
\providecommand \@url [1]{\endgroup\@href {#1}{\urlprefix }}%
\providecommand \urlprefix  [0]{URL }%
\providecommand \Eprint [0]{\href }%
\providecommand \doibase [0]{http://dx.doi.org/}%
\providecommand \selectlanguage [0]{\@gobble}%
\providecommand \bibinfo  [0]{\@secondoftwo}%
\providecommand \bibfield  [0]{\@secondoftwo}%
\providecommand \translation [1]{[#1]}%
\providecommand \BibitemOpen [0]{}%
\providecommand \bibitemStop [0]{}%
\providecommand \bibitemNoStop [0]{.\EOS\space}%
\providecommand \EOS [0]{\spacefactor3000\relax}%
\providecommand \BibitemShut  [1]{\csname bibitem#1\endcsname}%
\let\auto@bib@innerbib\@empty
%</preamble>
\bibitem [{\citenamefont {D'Innocenzo}\ \emph {et~al.}(2014)\citenamefont
  {D'Innocenzo}, \citenamefont {Srimath~Kandada}, \citenamefont {De~Bastiani},
  \citenamefont {Gandini},\ and\ \citenamefont {Petrozza}}]{DInnocenzo:2014aa}%
  \BibitemOpen
  \bibfield  {author} {\bibinfo {author} {\bibfnamefont {V.}~\bibnamefont
  {D'Innocenzo}}, \bibinfo {author} {\bibfnamefont {A.~R.}\ \bibnamefont
  {Srimath~Kandada}}, \bibinfo {author} {\bibfnamefont {M.}~\bibnamefont
  {De~Bastiani}}, \bibinfo {author} {\bibfnamefont {M.}~\bibnamefont
  {Gandini}}, \ and\ \bibinfo {author} {\bibfnamefont {A.}~\bibnamefont
  {Petrozza}},\ }\bibfield  {booktitle} {\emph {\bibinfo {booktitle} {Journal
  of the American Chemical Society}},\ }\href {\doibase 10.1021/ja511198f}
  {\bibfield  {journal} {\bibinfo  {journal} {Journal of the American Chemical
  Society}\ }\textbf {\bibinfo {volume} {136}},\ \bibinfo {pages} {17730}
  (\bibinfo {year} {2014})}\BibitemShut {NoStop}%
\bibitem [{\citenamefont {Kalantari}\ \emph {et~al.}(2011)\citenamefont
  {Kalantari}, \citenamefont {Sterianou}, \citenamefont {Karimi}, \citenamefont
  {Ferrarelli}, \citenamefont {Miao}, \citenamefont {Sinclair},\ and\
  \citenamefont {Reaney}}]{Kalantari2011}%
  \BibitemOpen
  \bibfield  {author} {\bibinfo {author} {\bibfnamefont {K.}~\bibnamefont
  {Kalantari}}, \bibinfo {author} {\bibfnamefont {I.}~\bibnamefont
  {Sterianou}}, \bibinfo {author} {\bibfnamefont {S.}~\bibnamefont {Karimi}},
  \bibinfo {author} {\bibfnamefont {M.~C.}\ \bibnamefont {Ferrarelli}},
  \bibinfo {author} {\bibfnamefont {S.}~\bibnamefont {Miao}}, \bibinfo {author}
  {\bibfnamefont {D.~C.}\ \bibnamefont {Sinclair}}, \ and\ \bibinfo {author}
  {\bibfnamefont {I.~M.}\ \bibnamefont {Reaney}},\ }\href {\doibase
  10.1002/adfm.201100191} {\bibfield  {journal} {\bibinfo  {journal} {Advanced
  Functional Materials}\ }\textbf {\bibinfo {volume} {21}},\ \bibinfo {pages}
  {3737} (\bibinfo {year} {2011})}\BibitemShut {NoStop}%
\bibitem [{\citenamefont {Gong}\ \emph {et~al.}(2014)\citenamefont {Gong},
  \citenamefont {Liu}, \citenamefont {Lupini}, \citenamefont {Shi},
  \citenamefont {Lin}, \citenamefont {Najmaei}, \citenamefont {Lin},
  \citenamefont {El{\'\i}as}, \citenamefont {Berkdemir}, \citenamefont {You},
  \citenamefont {Terrones}, \citenamefont {Terrones}, \citenamefont {Vajtai},
  \citenamefont {Pantelides}, \citenamefont {Pennycook}, \citenamefont {Lou},
  \citenamefont {Zhou},\ and\ \citenamefont {Ajayan}}]{Gong:2014aa}%
  \BibitemOpen
  \bibfield  {author} {\bibinfo {author} {\bibfnamefont {Y.}~\bibnamefont
  {Gong}}, \bibinfo {author} {\bibfnamefont {Z.}~\bibnamefont {Liu}}, \bibinfo
  {author} {\bibfnamefont {A.~R.}\ \bibnamefont {Lupini}}, \bibinfo {author}
  {\bibfnamefont {G.}~\bibnamefont {Shi}}, \bibinfo {author} {\bibfnamefont
  {J.}~\bibnamefont {Lin}}, \bibinfo {author} {\bibfnamefont {S.}~\bibnamefont
  {Najmaei}}, \bibinfo {author} {\bibfnamefont {Z.}~\bibnamefont {Lin}},
  \bibinfo {author} {\bibfnamefont {A.~L.}\ \bibnamefont {El{\'\i}as}},
  \bibinfo {author} {\bibfnamefont {A.}~\bibnamefont {Berkdemir}}, \bibinfo
  {author} {\bibfnamefont {G.}~\bibnamefont {You}}, \bibinfo {author}
  {\bibfnamefont {H.}~\bibnamefont {Terrones}}, \bibinfo {author}
  {\bibfnamefont {M.}~\bibnamefont {Terrones}}, \bibinfo {author}
  {\bibfnamefont {R.}~\bibnamefont {Vajtai}}, \bibinfo {author} {\bibfnamefont
  {S.~T.}\ \bibnamefont {Pantelides}}, \bibinfo {author} {\bibfnamefont
  {S.~J.}\ \bibnamefont {Pennycook}}, \bibinfo {author} {\bibfnamefont
  {J.}~\bibnamefont {Lou}}, \bibinfo {author} {\bibfnamefont {W.}~\bibnamefont
  {Zhou}}, \ and\ \bibinfo {author} {\bibfnamefont {P.~M.}\ \bibnamefont
  {Ajayan}},\ }\bibfield  {booktitle} {\emph {\bibinfo {booktitle} {Nano
  Letters}},\ }\href {\doibase 10.1021/nl4032296} {\bibfield  {journal}
  {\bibinfo  {journal} {Nano Letters}\ }\textbf {\bibinfo {volume} {14}},\
  \bibinfo {pages} {442} (\bibinfo {year} {2014})}\BibitemShut {NoStop}%
\bibitem [{\citenamefont {Jain}\ \emph {et~al.}(2013)\citenamefont {Jain},
  \citenamefont {Ong}, \citenamefont {Hautier}, \citenamefont {Chen},
  \citenamefont {Richards}, \citenamefont {Dacek}, \citenamefont {Cholia},
  \citenamefont {Gunter}, \citenamefont {Skinner}, \citenamefont {Ceder},\ and\
  \citenamefont {Persson}}]{MatProj2013}%
  \BibitemOpen
  \bibfield  {author} {\bibinfo {author} {\bibfnamefont {A.}~\bibnamefont
  {Jain}}, \bibinfo {author} {\bibfnamefont {S.~P.}\ \bibnamefont {Ong}},
  \bibinfo {author} {\bibfnamefont {G.}~\bibnamefont {Hautier}}, \bibinfo
  {author} {\bibfnamefont {W.}~\bibnamefont {Chen}}, \bibinfo {author}
  {\bibfnamefont {W.~D.}\ \bibnamefont {Richards}}, \bibinfo {author}
  {\bibfnamefont {S.}~\bibnamefont {Dacek}}, \bibinfo {author} {\bibfnamefont
  {S.}~\bibnamefont {Cholia}}, \bibinfo {author} {\bibfnamefont
  {D.}~\bibnamefont {Gunter}}, \bibinfo {author} {\bibfnamefont
  {D.}~\bibnamefont {Skinner}}, \bibinfo {author} {\bibfnamefont
  {G.}~\bibnamefont {Ceder}}, \ and\ \bibinfo {author} {\bibfnamefont {K.~A.}\
  \bibnamefont {Persson}},\ }\href {\doibase 10.1063/1.4812323} {\bibfield
  {journal} {\bibinfo  {journal} {APL Materials}\ }\textbf {\bibinfo {volume}
  {1}},\ \bibinfo {pages} {011002} (\bibinfo {year} {2013})},\ \Eprint
  {http://arxiv.org/abs/https://doi.org/10.1063/1.4812323}
  {https://doi.org/10.1063/1.4812323} \BibitemShut {NoStop}%
\bibitem [{\citenamefont {Saal}\ \emph {et~al.}(2013)\citenamefont {Saal},
  \citenamefont {Kirklin}, \citenamefont {Aykol}, \citenamefont {Meredig},\
  and\ \citenamefont {Wolverton}}]{OQMD}%
  \BibitemOpen
  \bibfield  {author} {\bibinfo {author} {\bibfnamefont {J.}~\bibnamefont
  {Saal}}, \bibinfo {author} {\bibfnamefont {S.}~\bibnamefont {Kirklin}},
  \bibinfo {author} {\bibfnamefont {M.}~\bibnamefont {Aykol}}, \bibinfo
  {author} {\bibfnamefont {B.}~\bibnamefont {Meredig}}, \ and\ \bibinfo
  {author} {\bibfnamefont {C.}~\bibnamefont {Wolverton}},\ }\href {\doibase
  10.1007/s11837-013-0755-4} {\bibfield  {journal} {\bibinfo  {journal} {JOM}\
  }\textbf {\bibinfo {volume} {65}},\ \bibinfo {pages} {1501} (\bibinfo {year}
  {2013})}\BibitemShut {NoStop}%
\bibitem [{\citenamefont {Jansen}(2015)}]{Jansen:2015aa}%
  \BibitemOpen
  \bibfield  {author} {\bibinfo {author} {\bibfnamefont {M.}~\bibnamefont
  {Jansen}},\ }\bibfield  {booktitle} {\emph {\bibinfo {booktitle} {Advanced
  Materials}},\ }\href {\doibase 10.1002/adma.201500143} {\bibfield  {journal}
  {\bibinfo  {journal} {Advanced Materials}\ }\textbf {\bibinfo {volume}
  {27}},\ \bibinfo {pages} {3229} (\bibinfo {year} {2015})}\BibitemShut
  {NoStop}%
\bibitem [{\citenamefont {Ludwig}(2019)}]{Ludwig:2019aa}%
  \BibitemOpen
  \bibfield  {author} {\bibinfo {author} {\bibfnamefont {A.}~\bibnamefont
  {Ludwig}},\ }\href {\doibase 10.1038/s41524-019-0205-0} {\bibfield  {journal}
  {\bibinfo  {journal} {npj Computational Materials}\ }\textbf {\bibinfo
  {volume} {5}},\ \bibinfo {pages} {70} (\bibinfo {year} {2019})}\BibitemShut
  {NoStop}%
\bibitem [{\citenamefont {Kanatzidis}(2017)}]{Kanatzidis:2017aa}%
  \BibitemOpen
  \bibfield  {author} {\bibinfo {author} {\bibfnamefont {M.~G.}\ \bibnamefont
  {Kanatzidis}},\ }\bibfield  {booktitle} {\emph {\bibinfo {booktitle}
  {Inorganic Chemistry}},\ }\href {\doibase 10.1021/acs.inorgchem.7b00188}
  {\bibfield  {journal} {\bibinfo  {journal} {Inorganic Chemistry}\ }\textbf
  {\bibinfo {volume} {56}},\ \bibinfo {pages} {3158} (\bibinfo {year}
  {2017})}\BibitemShut {NoStop}%
\bibitem [{\citenamefont {Aykol}\ \emph {et~al.}(2019)\citenamefont {Aykol},
  \citenamefont {Hegde}, \citenamefont {Hung}, \citenamefont {Suram},
  \citenamefont {Herring}, \citenamefont {Wolverton},\ and\ \citenamefont
  {Hummelsh{\o}j}}]{Aykol:2019aa}%
  \BibitemOpen
  \bibfield  {author} {\bibinfo {author} {\bibfnamefont {M.}~\bibnamefont
  {Aykol}}, \bibinfo {author} {\bibfnamefont {V.~I.}\ \bibnamefont {Hegde}},
  \bibinfo {author} {\bibfnamefont {L.}~\bibnamefont {Hung}}, \bibinfo {author}
  {\bibfnamefont {S.}~\bibnamefont {Suram}}, \bibinfo {author} {\bibfnamefont
  {P.}~\bibnamefont {Herring}}, \bibinfo {author} {\bibfnamefont
  {C.}~\bibnamefont {Wolverton}}, \ and\ \bibinfo {author} {\bibfnamefont
  {J.~S.}\ \bibnamefont {Hummelsh{\o}j}},\ }\href {\doibase
  10.1038/s41467-019-10030-5} {\bibfield  {journal} {\bibinfo  {journal}
  {Nature Communications}\ }\textbf {\bibinfo {volume} {10}},\ \bibinfo {pages}
  {2018} (\bibinfo {year} {2019})}\BibitemShut {NoStop}%
\bibitem [{\citenamefont {Kim}\ \emph {et~al.}(2019)\citenamefont {Kim},
  \citenamefont {Kim}, \citenamefont {Antono}, \citenamefont {Meredig},\ and\
  \citenamefont {Ling}}]{kim2019machinelearned}%
  \BibitemOpen
  \bibfield  {author} {\bibinfo {author} {\bibfnamefont {Y.}~\bibnamefont
  {Kim}}, \bibinfo {author} {\bibfnamefont {E.}~\bibnamefont {Kim}}, \bibinfo
  {author} {\bibfnamefont {E.}~\bibnamefont {Antono}}, \bibinfo {author}
  {\bibfnamefont {B.}~\bibnamefont {Meredig}}, \ and\ \bibinfo {author}
  {\bibfnamefont {J.}~\bibnamefont {Ling}},\ }\href@noop {} {\enquote {\bibinfo
  {title} {Machine-learned metrics for predicting the likelihood of success in
  materials discovery},}\ } (\bibinfo {year} {2019}),\ \Eprint
  {http://arxiv.org/abs/1911.11201} {arXiv:1911.11201 [cond-mat.mtrl-sci]}
  \BibitemShut {NoStop}%
\bibitem [{\citenamefont {Kononova}\ \emph {et~al.}(2019)\citenamefont
  {Kononova}, \citenamefont {Huo}, \citenamefont {He}, \citenamefont {Rong},
  \citenamefont {Botari}, \citenamefont {Sun}, \citenamefont {Tshitoyan},\ and\
  \citenamefont {Ceder}}]{Kononova:2019aa}%
  \BibitemOpen
  \bibfield  {author} {\bibinfo {author} {\bibfnamefont {O.}~\bibnamefont
  {Kononova}}, \bibinfo {author} {\bibfnamefont {H.}~\bibnamefont {Huo}},
  \bibinfo {author} {\bibfnamefont {T.}~\bibnamefont {He}}, \bibinfo {author}
  {\bibfnamefont {Z.}~\bibnamefont {Rong}}, \bibinfo {author} {\bibfnamefont
  {T.}~\bibnamefont {Botari}}, \bibinfo {author} {\bibfnamefont
  {W.}~\bibnamefont {Sun}}, \bibinfo {author} {\bibfnamefont {V.}~\bibnamefont
  {Tshitoyan}}, \ and\ \bibinfo {author} {\bibfnamefont {G.}~\bibnamefont
  {Ceder}},\ }\href {\doibase 10.1038/s41597-019-0224-1} {\bibfield  {journal}
  {\bibinfo  {journal} {Scientific Data}\ }\textbf {\bibinfo {volume} {6}},\
  \bibinfo {pages} {203} (\bibinfo {year} {2019})}\BibitemShut {NoStop}%
\bibitem [{\citenamefont {Kim}\ \emph {et~al.}(2020)\citenamefont {Kim},
  \citenamefont {Jensen}, \citenamefont {van Grootel}, \citenamefont {Huang},
  \citenamefont {Staib}, \citenamefont {Mysore}, \citenamefont {Chang},
  \citenamefont {Strubell}, \citenamefont {McCallum}, \citenamefont {Jegelka},\
  and\ \citenamefont {Olivetti}}]{kim2018inorganic}%
  \BibitemOpen
  \bibfield  {author} {\bibinfo {author} {\bibfnamefont {E.}~\bibnamefont
  {Kim}}, \bibinfo {author} {\bibfnamefont {Z.}~\bibnamefont {Jensen}},
  \bibinfo {author} {\bibfnamefont {A.}~\bibnamefont {van Grootel}}, \bibinfo
  {author} {\bibfnamefont {K.}~\bibnamefont {Huang}}, \bibinfo {author}
  {\bibfnamefont {M.}~\bibnamefont {Staib}}, \bibinfo {author} {\bibfnamefont
  {S.}~\bibnamefont {Mysore}}, \bibinfo {author} {\bibfnamefont {H.-S.}\
  \bibnamefont {Chang}}, \bibinfo {author} {\bibfnamefont {E.}~\bibnamefont
  {Strubell}}, \bibinfo {author} {\bibfnamefont {A.}~\bibnamefont {McCallum}},
  \bibinfo {author} {\bibfnamefont {S.}~\bibnamefont {Jegelka}}, \ and\
  \bibinfo {author} {\bibfnamefont {E.}~\bibnamefont {Olivetti}},\ }\href
  {\doibase 10.1021/acs.jcim.9b00995} {\bibfield  {journal} {\bibinfo
  {journal} {Journal of Chemical Information and Modeling}\ }\textbf {\bibinfo
  {volume} {60}},\ \bibinfo {pages} {1194} (\bibinfo {year} {2020})},\ \bibinfo
  {note} {pMID: 31909619},\ \Eprint
  {http://arxiv.org/abs/https://doi.org/10.1021/acs.jcim.9b00995}
  {https://doi.org/10.1021/acs.jcim.9b00995} \BibitemShut {NoStop}%
\bibitem [{\citenamefont {Ward}\ \emph {et~al.}(2016)\citenamefont {Ward},
  \citenamefont {Agrawal}, \citenamefont {Choudhary},\ and\ \citenamefont
  {Wolverton}}]{Ward:2016aa}%
  \BibitemOpen
  \bibfield  {author} {\bibinfo {author} {\bibfnamefont {L.}~\bibnamefont
  {Ward}}, \bibinfo {author} {\bibfnamefont {A.}~\bibnamefont {Agrawal}},
  \bibinfo {author} {\bibfnamefont {A.}~\bibnamefont {Choudhary}}, \ and\
  \bibinfo {author} {\bibfnamefont {C.}~\bibnamefont {Wolverton}},\ }\href
  {\doibase 10.1038/npjcompumats.2016.28} {\bibfield  {journal} {\bibinfo
  {journal} {npj Computational Materials}\ }\textbf {\bibinfo {volume} {2}},\
  \bibinfo {pages} {16028} (\bibinfo {year} {2016})}\BibitemShut {NoStop}%
\bibitem [{\citenamefont {Tshitoyan}\ \emph {et~al.}(2019)\citenamefont
  {Tshitoyan}, \citenamefont {Dagdelen}, \citenamefont {Weston}, \citenamefont
  {Dunn}, \citenamefont {Rong}, \citenamefont {Kononova}, \citenamefont
  {Persson}, \citenamefont {Ceder},\ and\ \citenamefont
  {Jain}}]{Tshitoyan:2019aa}%
  \BibitemOpen
  \bibfield  {author} {\bibinfo {author} {\bibfnamefont {V.}~\bibnamefont
  {Tshitoyan}}, \bibinfo {author} {\bibfnamefont {J.}~\bibnamefont {Dagdelen}},
  \bibinfo {author} {\bibfnamefont {L.}~\bibnamefont {Weston}}, \bibinfo
  {author} {\bibfnamefont {A.}~\bibnamefont {Dunn}}, \bibinfo {author}
  {\bibfnamefont {Z.}~\bibnamefont {Rong}}, \bibinfo {author} {\bibfnamefont
  {O.}~\bibnamefont {Kononova}}, \bibinfo {author} {\bibfnamefont {K.~A.}\
  \bibnamefont {Persson}}, \bibinfo {author} {\bibfnamefont {G.}~\bibnamefont
  {Ceder}}, \ and\ \bibinfo {author} {\bibfnamefont {A.}~\bibnamefont {Jain}},\
  }\href {\doibase 10.1038/s41586-019-1335-8} {\bibfield  {journal} {\bibinfo
  {journal} {Nature}\ }\textbf {\bibinfo {volume} {571}},\ \bibinfo {pages}
  {95} (\bibinfo {year} {2019})}\BibitemShut {NoStop}%
\bibitem [{\citenamefont {Sutskever}\ \emph {et~al.}(2014)\citenamefont
  {Sutskever}, \citenamefont {Vinyals},\ and\ \citenamefont {Le}}]{seq2seq}%
  \BibitemOpen
  \bibfield  {author} {\bibinfo {author} {\bibfnamefont {I.}~\bibnamefont
  {Sutskever}}, \bibinfo {author} {\bibfnamefont {O.}~\bibnamefont {Vinyals}},
  \ and\ \bibinfo {author} {\bibfnamefont {Q.~V.}\ \bibnamefont {Le}},\ }\href
  {http://arxiv.org/abs/1409.3215} {\bibfield  {journal} {\bibinfo  {journal}
  {CoRR}\ }\textbf {\bibinfo {volume} {abs/1409.3215}} (\bibinfo {year}
  {2014})},\ \Eprint {http://arxiv.org/abs/1409.3215} {arXiv:1409.3215}
  \BibitemShut {NoStop}%
\bibitem [{\citenamefont {Vaswani}\ \emph {et~al.}(2017)\citenamefont
  {Vaswani}, \citenamefont {Shazeer}, \citenamefont {Parmar}, \citenamefont
  {Uszkoreit}, \citenamefont {Jones}, \citenamefont {Gomez}, \citenamefont
  {Kaiser},\ and\ \citenamefont {Polosukhin}}]{vaswani2017attention}%
  \BibitemOpen
  \bibfield  {author} {\bibinfo {author} {\bibfnamefont {A.}~\bibnamefont
  {Vaswani}}, \bibinfo {author} {\bibfnamefont {N.}~\bibnamefont {Shazeer}},
  \bibinfo {author} {\bibfnamefont {N.}~\bibnamefont {Parmar}}, \bibinfo
  {author} {\bibfnamefont {J.}~\bibnamefont {Uszkoreit}}, \bibinfo {author}
  {\bibfnamefont {L.}~\bibnamefont {Jones}}, \bibinfo {author} {\bibfnamefont
  {A.~N.}\ \bibnamefont {Gomez}}, \bibinfo {author} {\bibfnamefont
  {L.}~\bibnamefont {Kaiser}}, \ and\ \bibinfo {author} {\bibfnamefont
  {I.}~\bibnamefont {Polosukhin}},\ }\href@noop {} {\enquote {\bibinfo {title}
  {Attention is all you need},}\ } (\bibinfo {year} {2017}),\ \Eprint
  {http://arxiv.org/abs/1706.03762} {arXiv:1706.03762 [cs.CL]} \BibitemShut
  {NoStop}%
\bibitem [{\citenamefont {Devlin}\ \emph {et~al.}(2018)\citenamefont {Devlin},
  \citenamefont {Chang}, \citenamefont {Lee},\ and\ \citenamefont
  {Toutanova}}]{devlin2018bert}%
  \BibitemOpen
  \bibfield  {author} {\bibinfo {author} {\bibfnamefont {J.}~\bibnamefont
  {Devlin}}, \bibinfo {author} {\bibfnamefont {M.-W.}\ \bibnamefont {Chang}},
  \bibinfo {author} {\bibfnamefont {K.}~\bibnamefont {Lee}}, \ and\ \bibinfo
  {author} {\bibfnamefont {K.}~\bibnamefont {Toutanova}},\ }\href@noop {}
  {\enquote {\bibinfo {title} {Bert: Pre-training of deep bidirectional
  transformers for language understanding},}\ } (\bibinfo {year} {2018}),\
  \Eprint {http://arxiv.org/abs/1810.04805} {arXiv:1810.04805 [cs.CL]}
  \BibitemShut {NoStop}%
\bibitem [{\citenamefont {Hochreiter}\ and\ \citenamefont
  {Schmidhuber}(1997)}]{lstm}%
  \BibitemOpen
  \bibfield  {author} {\bibinfo {author} {\bibfnamefont {S.}~\bibnamefont
  {Hochreiter}}\ and\ \bibinfo {author} {\bibfnamefont {J.}~\bibnamefont
  {Schmidhuber}},\ }\href {\doibase 10.1162/neco.1997.9.8.1735} {\bibfield
  {journal} {\bibinfo  {journal} {Neural Comput.}\ }\textbf {\bibinfo {volume}
  {9}},\ \bibinfo {pages} {1735} (\bibinfo {year} {1997})}\BibitemShut
  {NoStop}%
\bibitem [{\citenamefont {Goodall}\ and\ \citenamefont
  {Lee}(2019)}]{goodall2019predicting}%
  \BibitemOpen
  \bibfield  {author} {\bibinfo {author} {\bibfnamefont {R.~E.~A.}\
  \bibnamefont {Goodall}}\ and\ \bibinfo {author} {\bibfnamefont {A.~A.}\
  \bibnamefont {Lee}},\ }\href@noop {} {\enquote {\bibinfo {title} {Predicting
  materials properties without crystal structure: Deep representation learning
  from stoichiometry},}\ } (\bibinfo {year} {2019}),\ \Eprint
  {http://arxiv.org/abs/1910.00617} {arXiv:1910.00617 [physics.comp-ph]}
  \BibitemShut {NoStop}%
\bibitem [{\citenamefont {Veli{\v c}kovi{\'c}}\ \emph
  {et~al.}(2017)\citenamefont {Veli{\v c}kovi{\'c}}, \citenamefont {Cucurull},
  \citenamefont {Casanova}, \citenamefont {Romero}, \citenamefont {Li{\`o}},\
  and\ \citenamefont {Bengio}}]{velikovi2017graph}%
  \BibitemOpen
  \bibfield  {author} {\bibinfo {author} {\bibfnamefont {P.}~\bibnamefont
  {Veli{\v c}kovi{\'c}}}, \bibinfo {author} {\bibfnamefont {G.}~\bibnamefont
  {Cucurull}}, \bibinfo {author} {\bibfnamefont {A.}~\bibnamefont {Casanova}},
  \bibinfo {author} {\bibfnamefont {A.}~\bibnamefont {Romero}}, \bibinfo
  {author} {\bibfnamefont {P.}~\bibnamefont {Li{\`o}}}, \ and\ \bibinfo
  {author} {\bibfnamefont {Y.}~\bibnamefont {Bengio}},\ }\href@noop {}
  {\enquote {\bibinfo {title} {Graph attention networks},}\ } (\bibinfo {year}
  {2017}),\ \Eprint {http://arxiv.org/abs/1710.10903} {arXiv:1710.10903
  [stat.ML]} \BibitemShut {NoStop}%
\bibitem [{\citenamefont {Gover}\ \emph {et~al.}(2006)\citenamefont {Gover},
  \citenamefont {Bryan}, \citenamefont {Burns},\ and\ \citenamefont
  {Barker}}]{GOVER20061495}%
  \BibitemOpen
  \bibfield  {author} {\bibinfo {author} {\bibfnamefont {R.}~\bibnamefont
  {Gover}}, \bibinfo {author} {\bibfnamefont {A.}~\bibnamefont {Bryan}},
  \bibinfo {author} {\bibfnamefont {P.}~\bibnamefont {Burns}}, \ and\ \bibinfo
  {author} {\bibfnamefont {J.}~\bibnamefont {Barker}},\ }\href {\doibase
  https://doi.org/10.1016/j.ssi.2006.07.028} {\bibfield  {journal} {\bibinfo
  {journal} {Solid State Ionics}\ }\textbf {\bibinfo {volume} {177}},\ \bibinfo
  {pages} {1495 } (\bibinfo {year} {2006})}\BibitemShut {NoStop}%
\bibitem [{\citenamefont {Tokura}\ \emph {et~al.}(1994)\citenamefont {Tokura},
  \citenamefont {Urushibara}, \citenamefont {Moritomo}, \citenamefont {Arima},
  \citenamefont {Asamitsu}, \citenamefont {Kido},\ and\ \citenamefont
  {Furukawa}}]{magnetores}%
  \BibitemOpen
  \bibfield  {author} {\bibinfo {author} {\bibfnamefont {Y.}~\bibnamefont
  {Tokura}}, \bibinfo {author} {\bibfnamefont {A.}~\bibnamefont {Urushibara}},
  \bibinfo {author} {\bibfnamefont {Y.}~\bibnamefont {Moritomo}}, \bibinfo
  {author} {\bibfnamefont {T.}~\bibnamefont {Arima}}, \bibinfo {author}
  {\bibfnamefont {A.}~\bibnamefont {Asamitsu}}, \bibinfo {author}
  {\bibfnamefont {G.}~\bibnamefont {Kido}}, \ and\ \bibinfo {author}
  {\bibfnamefont {N.}~\bibnamefont {Furukawa}},\ }\href {\doibase
  10.1143/JPSJ.63.3931} {\bibfield  {journal} {\bibinfo  {journal} {Journal of
  the Physical Society of Japan}\ }\textbf {\bibinfo {volume} {63}},\ \bibinfo
  {pages} {3931} (\bibinfo {year} {1994})},\ \Eprint
  {http://arxiv.org/abs/https://doi.org/10.1143/JPSJ.63.3931}
  {https://doi.org/10.1143/JPSJ.63.3931} \BibitemShut {NoStop}%
\bibitem [{\citenamefont {Debbebi}\ \emph {et~al.}(2018)\citenamefont
  {Debbebi}, \citenamefont {Omrani}, \citenamefont {Cheikhrouhou-Koubaa},\ and\
  \citenamefont {Cheikhrouhou}}]{DEBBEBI201867}%
  \BibitemOpen
  \bibfield  {author} {\bibinfo {author} {\bibfnamefont {I.~S.}\ \bibnamefont
  {Debbebi}}, \bibinfo {author} {\bibfnamefont {H.}~\bibnamefont {Omrani}},
  \bibinfo {author} {\bibfnamefont {W.}~\bibnamefont {Cheikhrouhou-Koubaa}}, \
  and\ \bibinfo {author} {\bibfnamefont {A.}~\bibnamefont {Cheikhrouhou}},\
  }\href {\doibase https://doi.org/10.1016/j.jpcs.2017.10.018} {\bibfield
  {journal} {\bibinfo  {journal} {Journal of Physics and Chemistry of Solids}\
  }\textbf {\bibinfo {volume} {113}},\ \bibinfo {pages} {67 } (\bibinfo {year}
  {2018})}\BibitemShut {NoStop}%
\bibitem [{\citenamefont {Zhou}\ \emph {et~al.}(2006)\citenamefont {Zhou},
  \citenamefont {Zhang}, \citenamefont {Xie}, \citenamefont {Xiao},
  \citenamefont {Yang}, \citenamefont {He},\ and\ \citenamefont
  {Zhao}}]{ZHOU2006338}%
  \BibitemOpen
  \bibfield  {author} {\bibinfo {author} {\bibfnamefont {H.}~\bibnamefont
  {Zhou}}, \bibinfo {author} {\bibfnamefont {X.}~\bibnamefont {Zhang}},
  \bibinfo {author} {\bibfnamefont {B.}~\bibnamefont {Xie}}, \bibinfo {author}
  {\bibfnamefont {Y.}~\bibnamefont {Xiao}}, \bibinfo {author} {\bibfnamefont
  {C.}~\bibnamefont {Yang}}, \bibinfo {author} {\bibfnamefont {Y.}~\bibnamefont
  {He}}, \ and\ \bibinfo {author} {\bibfnamefont {Y.}~\bibnamefont {Zhao}},\
  }\href {\doibase https://doi.org/10.1016/j.tsf.2005.10.078} {\bibfield
  {journal} {\bibinfo  {journal} {Thin Solid Films}\ }\textbf {\bibinfo
  {volume} {497}},\ \bibinfo {pages} {338 } (\bibinfo {year}
  {2006})}\BibitemShut {NoStop}%
\bibitem [{\citenamefont {Raccuglia}\ \emph {et~al.}(2016)\citenamefont
  {Raccuglia}, \citenamefont {Elbert}, \citenamefont {Adler}, \citenamefont
  {Falk}, \citenamefont {Wenny}, \citenamefont {Mollo}, \citenamefont {Zeller},
  \citenamefont {Friedler}, \citenamefont {Schrier},\ and\ \citenamefont
  {Norquist}}]{Raccuglia:2016aa}%
  \BibitemOpen
  \bibfield  {author} {\bibinfo {author} {\bibfnamefont {P.}~\bibnamefont
  {Raccuglia}}, \bibinfo {author} {\bibfnamefont {K.~C.}\ \bibnamefont
  {Elbert}}, \bibinfo {author} {\bibfnamefont {P.~D.~F.}\ \bibnamefont
  {Adler}}, \bibinfo {author} {\bibfnamefont {C.}~\bibnamefont {Falk}},
  \bibinfo {author} {\bibfnamefont {M.~B.}\ \bibnamefont {Wenny}}, \bibinfo
  {author} {\bibfnamefont {A.}~\bibnamefont {Mollo}}, \bibinfo {author}
  {\bibfnamefont {M.}~\bibnamefont {Zeller}}, \bibinfo {author} {\bibfnamefont
  {S.~A.}\ \bibnamefont {Friedler}}, \bibinfo {author} {\bibfnamefont
  {J.}~\bibnamefont {Schrier}}, \ and\ \bibinfo {author} {\bibfnamefont
  {A.~J.}\ \bibnamefont {Norquist}},\ }\href {\doibase 10.1038/nature17439}
  {\bibfield  {journal} {\bibinfo  {journal} {Nature}\ }\textbf {\bibinfo
  {volume} {533}},\ \bibinfo {pages} {73} (\bibinfo {year} {2016})}\BibitemShut
  {NoStop}%
\bibitem [{\citenamefont {Paszke}\ \emph {et~al.}(2019)\citenamefont {Paszke},
  \citenamefont {Gross}, \citenamefont {Massa}, \citenamefont {Lerer},
  \citenamefont {Bradbury}, \citenamefont {Chanan}, \citenamefont {Killeen},
  \citenamefont {Lin}, \citenamefont {Gimelshein}, \citenamefont {Antiga},
  \citenamefont {Desmaison}, \citenamefont {Kopf}, \citenamefont {Yang},
  \citenamefont {DeVito}, \citenamefont {Raison}, \citenamefont {Tejani},
  \citenamefont {Chilamkurthy}, \citenamefont {Steiner}, \citenamefont {Fang},
  \citenamefont {Bai},\ and\ \citenamefont {Chintala}}]{NEURIPS2019_9015}%
  \BibitemOpen
  \bibfield  {author} {\bibinfo {author} {\bibfnamefont {A.}~\bibnamefont
  {Paszke}}, \bibinfo {author} {\bibfnamefont {S.}~\bibnamefont {Gross}},
  \bibinfo {author} {\bibfnamefont {F.}~\bibnamefont {Massa}}, \bibinfo
  {author} {\bibfnamefont {A.}~\bibnamefont {Lerer}}, \bibinfo {author}
  {\bibfnamefont {J.}~\bibnamefont {Bradbury}}, \bibinfo {author}
  {\bibfnamefont {G.}~\bibnamefont {Chanan}}, \bibinfo {author} {\bibfnamefont
  {T.}~\bibnamefont {Killeen}}, \bibinfo {author} {\bibfnamefont
  {Z.}~\bibnamefont {Lin}}, \bibinfo {author} {\bibfnamefont {N.}~\bibnamefont
  {Gimelshein}}, \bibinfo {author} {\bibfnamefont {L.}~\bibnamefont {Antiga}},
  \bibinfo {author} {\bibfnamefont {A.}~\bibnamefont {Desmaison}}, \bibinfo
  {author} {\bibfnamefont {A.}~\bibnamefont {Kopf}}, \bibinfo {author}
  {\bibfnamefont {E.}~\bibnamefont {Yang}}, \bibinfo {author} {\bibfnamefont
  {Z.}~\bibnamefont {DeVito}}, \bibinfo {author} {\bibfnamefont
  {M.}~\bibnamefont {Raison}}, \bibinfo {author} {\bibfnamefont
  {A.}~\bibnamefont {Tejani}}, \bibinfo {author} {\bibfnamefont
  {S.}~\bibnamefont {Chilamkurthy}}, \bibinfo {author} {\bibfnamefont
  {B.}~\bibnamefont {Steiner}}, \bibinfo {author} {\bibfnamefont
  {L.}~\bibnamefont {Fang}}, \bibinfo {author} {\bibfnamefont {J.}~\bibnamefont
  {Bai}}, \ and\ \bibinfo {author} {\bibfnamefont {S.}~\bibnamefont
  {Chintala}},\ }in\ \href
  {http://papers.neurips.cc/paper/9015-pytorch-an-imperative-style-high-performance-deep-learning-library.pdf}
  {\emph {\bibinfo {booktitle} {Advances in Neural Information Processing
  Systems 32}}},\ \bibinfo {editor} {edited by\ \bibinfo {editor}
  {\bibfnamefont {H.}~\bibnamefont {Wallach}}, \bibinfo {editor} {\bibfnamefont
  {H.}~\bibnamefont {Larochelle}}, \bibinfo {editor} {\bibfnamefont
  {A.}~\bibnamefont {Beygelzimer}}, \bibinfo {editor} {\bibfnamefont
  {F.}~\bibnamefont {d'Alch\'{e} Buc}}, \bibinfo {editor} {\bibfnamefont
  {E.}~\bibnamefont {Fox}}, \ and\ \bibinfo {editor} {\bibfnamefont
  {R.}~\bibnamefont {Garnett}}}\ (\bibinfo  {publisher} {Curran Associates,
  Inc.},\ \bibinfo {year} {2019})\ pp.\ \bibinfo {pages}
  {8024--8035}\BibitemShut {NoStop}%
\bibitem [{\citenamefont {Ward}\ \emph {et~al.}(2018)\citenamefont {Ward},
  \citenamefont {Dunn}, \citenamefont {Faghaninia}, \citenamefont {Zimmermann},
  \citenamefont {Bajaj}, \citenamefont {Wang}, \citenamefont {Montoya},
  \citenamefont {Chen}, \citenamefont {Bystrom}, \citenamefont {Dylla},
  \citenamefont {Chard}, \citenamefont {Asta}, \citenamefont {Persson},
  \citenamefont {Snyder}, \citenamefont {Foster},\ and\ \citenamefont
  {Jain}}]{matminer}%
  \BibitemOpen
  \bibfield  {author} {\bibinfo {author} {\bibfnamefont {L.}~\bibnamefont
  {Ward}}, \bibinfo {author} {\bibfnamefont {A.}~\bibnamefont {Dunn}}, \bibinfo
  {author} {\bibfnamefont {A.}~\bibnamefont {Faghaninia}}, \bibinfo {author}
  {\bibfnamefont {N.}~\bibnamefont {Zimmermann}}, \bibinfo {author}
  {\bibfnamefont {S.}~\bibnamefont {Bajaj}}, \bibinfo {author} {\bibfnamefont
  {Q.}~\bibnamefont {Wang}}, \bibinfo {author} {\bibfnamefont {J.}~\bibnamefont
  {Montoya}}, \bibinfo {author} {\bibfnamefont {J.}~\bibnamefont {Chen}},
  \bibinfo {author} {\bibfnamefont {K.}~\bibnamefont {Bystrom}}, \bibinfo
  {author} {\bibfnamefont {M.}~\bibnamefont {Dylla}}, \bibinfo {author}
  {\bibfnamefont {K.}~\bibnamefont {Chard}}, \bibinfo {author} {\bibfnamefont
  {M.}~\bibnamefont {Asta}}, \bibinfo {author} {\bibfnamefont {K.}~\bibnamefont
  {Persson}}, \bibinfo {author} {\bibfnamefont {G.}~\bibnamefont {Snyder}},
  \bibinfo {author} {\bibfnamefont {I.}~\bibnamefont {Foster}}, \ and\ \bibinfo
  {author} {\bibfnamefont {A.}~\bibnamefont {Jain}},\ }\href {\doibase
  10.1016/j.commatsci.2018.05.018} {\bibfield  {journal} {\bibinfo  {journal}
  {Computational Materials Science}\ }\textbf {\bibinfo {volume} {152}},\
  \bibinfo {pages} {60} (\bibinfo {year} {2018})}\BibitemShut {NoStop}%
\bibitem [{\citenamefont {He}\ \emph {et~al.}(2016)\citenamefont {He},
  \citenamefont {Zhang}, \citenamefont {Ren},\ and\ \citenamefont
  {Sun}}]{resnet}%
  \BibitemOpen
  \bibfield  {author} {\bibinfo {author} {\bibfnamefont {K.}~\bibnamefont
  {He}}, \bibinfo {author} {\bibfnamefont {X.}~\bibnamefont {Zhang}}, \bibinfo
  {author} {\bibfnamefont {S.}~\bibnamefont {Ren}}, \ and\ \bibinfo {author}
  {\bibfnamefont {J.}~\bibnamefont {Sun}},\ }in\ \href@noop {} {\emph {\bibinfo
  {booktitle} {2016 IEEE Conference on Computer Vision and Pattern Recognition
  (CVPR)}}}\ (\bibinfo {year} {2016})\ pp.\ \bibinfo {pages}
  {770--778}\BibitemShut {NoStop}%
\bibitem [{\citenamefont {Kingma}\ and\ \citenamefont
  {Ba}(2014)}]{kingma2014adam}%
  \BibitemOpen
  \bibfield  {author} {\bibinfo {author} {\bibfnamefont {D.~P.}\ \bibnamefont
  {Kingma}}\ and\ \bibinfo {author} {\bibfnamefont {J.}~\bibnamefont {Ba}},\
  }\href@noop {} {\enquote {\bibinfo {title} {Adam: A method for stochastic
  optimization},}\ } (\bibinfo {year} {2014}),\ \Eprint
  {http://arxiv.org/abs/1412.6980} {arXiv:1412.6980 [cs.LG]} \BibitemShut
  {NoStop}%
\end{thebibliography}%

\end{document}